\documentclass[12pt]{article}
\begin{document}
\centerline{\Large\bf Temporal Platonic Metaphysics\footnote{En entry to the FQXi Inagural Essay Contest on the Nature of Time.}}

\bigskip
\bigskip
\bigskip
\centerline{\large Aleksandar Mikovi\'c\footnote{Member of the Mathematical Physics Group, University of Lisbon. E-mail: amikovic@ulusofona.pt}}
\begin{center}
Departamento de Matem\'atica,
Universidade Lus\'ofona, Av. do Campo Grande, 376, 1749-024,
Lisboa, Portugal
\end{center}

\bigskip
\bigskip
\bigskip
\begin{abstract}
In this essay we describe a platonic metaphysics where time is a fundamental idea such that the passage of time is  independent of observers and the laws of physics. Furthermore, time serves to distinguish between a real and an abstract universe, where a real universe is an appropriatelly ordered set of ideas in time, while an abstract universe is such a set outside of time. We allow for non-computable and non-mathematical ideas to be part of our model of the universe in order to include intelligent observers. The part of the universe which can be described by mathematics corresponds to a mathematical structure in Tegmark's sense, while the non-mathematical part can be described by a human language. We describe the mind-brain connection in our framework, and show that it resolves the problem of obtaining abstract knowledge in platonism. We briefly discuss how different interpretations of Quantum Mechanics can be implemented in our framework and point out some physics implications, most notably that the time travel would be impossible.
\end{abstract}

\section{Introduction}

Plato's idea that our ideas have existence independent of us and outside of space and time has been developed into a philosophical system known as platonism, see \cite{pl}. Recently, the platonic approach has been used by M. Tegmark to construct a theory of a mathematical universe \cite{teg}, where it was postulated that our universe is the same as the ultimate mathematical structure which describes it. Note that this is a reduced version of platonism, since in Tegmark's approach the world of ideas consists only of certain mathematical structures. The main assumption of Tegmark is that everything that exists is reducible to a finite mathematical structure, and furthermore, he restricts the world of ideas to computable mathematical structures. Another salient feature of Tegmark's approach is that time is not a fundamental concept, but a phenomenon experienced by an observer. 

These assumptions are acceptable if one believes that time is not fundamental and that the human mind is computable. In this essay we are going to advocate a different platonic metaphysics, namely the one where time is a fundamental concept. Furthermore, the time will serve to distinguish between an abstract and a real universe. We will also  allow for non-computable and non-mathematical ideas to be part of a model of our universe. The computable part will be then described by the fundamental physical laws, i.e. by a mathematical structure in Tegmark's sense. The non-computable part will then require another set of laws, so that psychology could not be derived from the Standard Model.

There are strong arguments that the human mind is not computable, based on Goedel's theorems in logic, see \cite{penr}. As far as time is concerned, the arguments which are usually invoked against it are: 
\begin{enumerate}
	\item The mathematical structure of General Relativity (GR) is that of a 4-manifold with a pseudo-Riemannian metric and considering a time-like coordinate as time does not give a unique definition of time. Also there is no unique foliation of the spacetime manifold into space-like hypersurfaces.
	\item In quantum gravity, finding a time variable in general case is difficult \cite{tvar}.
\end{enumerate}
Clearly these arguments are related to the problem of how to define time within GR or within quantum gravity, but a lack of a good mathematical definition of time does not imply that time is an illusion or an emergent property. 

There are various indications that there is something fundamental concerning time. If time were not a fundamental concept, then one would expect that the physical metrics would have Euclidean signatures so that there would be no special time-like directions. On the other hand, spacetime metrics with Minkowski signatures allow us to split the spacetime into space and time\footnote{This can be always done locally, while globally can be done for a large class of metrics, i.e. the globally hyperbolic ones.}. In quantum gravity the split into space and time is preserved at the topological level when the canonical quantization formalism is used. The topology $S\times\bf R$ with a preferred space-like foliation is also used in the discretized path-integral approach of Dynamical Causal Triangulations where de Sitter metric can be recovered in the semiclassical limit \cite{ajl}.

We will consider the clock time as a rate of passage of time. This rate will depend on the clock trajectory, so that there will be no conflict with Special Theory of Relativity (STR). However, the passage of time will be absolute, in the sense that it will be observer-independent phenomenon. We will also consider the passage of time as a fundamental concept, not reducible to other simpler concepts. We will implement this idea in a Platonic metaphysics and we will postulate that our Universe is a more general structure then the one used by Tegmark in order to allow for the ideas of time-like change and intelligence to be irreducible.

\section{The world of ideas}

We will consider that everything that exists, be it a real object in spacetime or an abstract idea, to be an idea.
We will call the set of all ideas the World of Ideas (WI), and let us divide WI into mathematical and non-mathematical ideas. 

\subsection{Mathematical ideas}

The mathematical ideas can be characterized by using finite sets (i.e. finite sets of integers or finite sets of symbols) and relations between them. This can be pictured as a collection of labeled graphs, and a mathematical idea or a mathematical structure (a collection of mathematical ideas) would be a (finite) graph whose vertices carry finite ordered sequences of integers. One can then introduce notions of equivalent graphs, subgraphs, reducible graphs, irreducible graphs, symmetries, etc, see \cite{teg}. 

The set of finite graphs with finite-length labels corresponds to the mathematical structures used by Tegmark. However, one can imagine more general structures, like graphs with infinite-length labels or infinite graphs.

In the set of labeled graphs $LG$ one can introduce relations, which means that we need to extend
$LG$ to 
$$RG = LG \cup (LG)^2 \cup (LG)^3 \cup \cdots \cup (LG)^n \cup \cdots\,.$$ 
Let us consider a subset $O$ of $RG$, i.e. a set of ideas with relations. Again one can think of these objects as labeled graphs, but now the vertices are elements of $LG$, i.e. the ideas. Let us define a history $H$ as a subset of $O$ such that 
$$ H = \bigcup_{t\in I} S_{t}\,,$$
where $I \subseteq\bf R$ and $S_{t} \prec S_{t'}$ if $t < t'$ and $\prec$ is a relation of order in $O$.

This definition is based on the model of a universe where the spacetime is a manifold of topology $S\times \bf R$, while the elementary particles are described by their world-lines, or the world-tubes, in the case of String Theory. If we take $I\subseteq \bf Z$, then we can consider $H$ as a generalization of the simplicial complex corresponding to the spacetime manifold containing  matter. The sets $S_t$ can be considered as generalizations of the spatial sections of $S\times \bf R$. The links (relations) between the elements of $S_t$ will be called spatial relations, while the links between the elements of $S_t$ and $S_{t'}$, where $t\ne t'$, will be called temporal relations.

\subsection{Time}

Let us introduce the idea of time-like change or passage of time $T$. We will assume that
$T$ is a non-mathematical and an irreducible idea. A temporal universe $U$ would be a $(H,T)$ pair, where the instant of now is moving along the levels $S_t$ of $H$ with $t$ increasing. This is based on our experience of time passage when observing a particle (or an object) changing its position in space, or when observing an object changing its form. The identity of a particle or an object will be determined by its temporal relations, i.e. by its world-line or a world-tube. A simplified picture of a temporal universe would be a $D$-dimensional manifold with a moving $(D-1)$-dimensional submanifold. 

This picture may seem to be at odds with Special Theory of Relativity (STR), but there is no conflict, because the time in STR refers to clock readings, which are trajectory dependent. The clock time can be viewed as a rate of passage of time, and that rate depends on a clock trajectory and the laws of physics, while the passage of time is absolute, in the sense that is independent from a clock trajectory and from the laws of physics.

Another analogy would be that a history $H$ can be considered as a digital video disc (DVD) and $T$ as a DVD-player, so that a temporal universe $U(H,T)$ would be the corresponding movie played on a screen. When comparing to Tegmark's approach, one can see that for him any history $H$ with appropriate features would be a universe like the one in which we live. For us a history $H$ would be a timeless universe, and only the composition $U(H,T)$ would be a universe like ours.

\subsection{The observers}

When constructing a theory of everything, or more generally a metaphysics, one has to define the intelligent beings or the observers. In Tegmark's approach the observes are part of the ultimate mathematical structure $MS$, which is the same as the history $H$. Since we do not believe that a human mind can be completely described by mathematics, we will assume that intelligence is a non-mathematical idea and therefore for a universe with observers we will have that $MS \subset H$.

We can think of an observer as a bound state of elementary particles, which can be pictured as a set of intricately braided trajectories, with a special structure, called brain. The brain has capability of capturing and storing ideas, which can be mathematical and non-mathematical. The mind at some instant of time will be the set of ideas stored in the brain, or more generally it can be considered as a temporal sequence of sets of ideas with corresponding relations stored in our brain. The ideas stored in our brain can be of two types: the ideas which are records of our interactions with the environment and the ideas which are not this, i.e. the abstract ideas. 

A simplified picture of an observer would be a single world-line with a finite set of hooks attached to it. The hooks would represent brain at distinct moments of time. The ideas in observer's brain could be then pictured as colored graphs attached to the hooks. The number of ideas on a hook is increasing with time\footnote{At short time scales the number of ideas in a brain could decrease due to memory loss, but this number increases or stays constant over the longer periods.}, corresponding to our increasing memory and knowledge. Preserving the identity of an observer requires that 
$$ I_1 \subseteq I_2 \subseteq \cdots \subseteq I_n \,,$$
where $I_k$ is the set of ideas in his brain at time instant $t_k$ and $n$ is his lifespan.

The process of forming a theory for a certain set of phenomena can be then explained in the following way. An observer will have the record of his interactions with the environment related to these phenomena, i.e. the results of experiments. Let us denote this set of ideas at the moment $t_1$ as $R_1$ . At the moment $t_2$ he will have in his brain a mathematical structure $M_2 = M$ which will be related to $R_2 = R_1$, i.e. which is explaining the experimental results. The new experimental results $R_3$ at the moment $t_3$ will be then correlated with the structure $M_3 = M$ so that at the moment $t_4$ he will either have $M_4 = M$, if the new results can be explained by $M$, or $M_4 = M'$, if the new results can be explained by another mathematical structure $M'$, and so on. 

Since we do not believe in reductionism, then there will be phenomena where the observer cannot use only a mathematical structure as his theory. Such a theory will be then described in terms of a human language (this is situation in biology, psychology, social sciences and philosophy). A human language looks like a mathematical structure (finite list of symbols with finite length words and sentences) but the meaning or truthfulness of words and sentences, when they do not refer to a mathematical structure, is not established by relating them to a finite fixed set of axioms. The meaning is established by relating them to the world around us (which is reflected in our minds).

\subsection{Our universe}

The following simplified description of our universe can be now given. We can imagine our universe as the simplicial complex corresponding to a spacetime manifold of dimension $D$, such that the vertices contain labeled graphs, i.e. ideas. There will be a preferred foliation of the complex into subcomplexes of dimension $D-1$, and the instant of now will run through these subcomplexes. We will take $t\ge 0$, corresponding to a universe with a beginning. At early times the vertices will contain only the $n$-tuples of numbers, corresponding to the fundamental constituents of space and matter. At some later time the hooks (brains) will appear and the vertices will also contain graphs (ideas) corresponding to minds of intelligent beings.  

The edges of the spacetime simplicial complex would correspond to the temporal and the spatial relations of the history $H$ of our universe\footnote{For a realistic $H$ the corresponding graph will be more complicated than the one-complex of a $D$-dimensional manifold.}. The physical part of the universe will be given by the skeleton of $H$, $MS(H)$, which is the largest subgraph of $H$ given by a mathematical structure. However, the observers will contain non-mathematical ideas beside the mathematical ones, and this by definition can not be a part of some mathematical structure.

A useful analogy would be that of a decorated tree, where the tree is a mathematical structure, while the decorations are the minds of intelligent beings. The decorations are not the part of the tree mathematical structure. This decorated tree would represent a timeless history of a universe with observers, and in order to turn it into a living universe, we need to ``switch on'' the time, i.e. to compose it with the idea of passage of time $T$. A temporal universe would be then analogous to a burning decorated tree, where the flames would form a transversal surface and move from the bottom to the top of the tree. The burned part of the tree would be the past, the burning part would be the present and the future would be the part above the flames.

The universe we have described so far is essentially a classical one, but our approach is sufficiently general so that a quantum universe can be accommodated. How to implement this will depend on the interpretational scheme of Quantum Mechanics (QM) one believes in. However, all interpretations share a common feature: one needs multiple universes in order to accommodate the probabilistic nature of QM, as well as the existence of states which are linear combinations of two or more classically different states. 

For example, the Many Worlds interpretation, for a review and references see \cite{mwi}, will require history branchings, i.e. for $t \ge t_0$ the sets $S_t$ in $H$ will be disjoint unions of sets $S^{(k)}_t$ such that there are no relations among them, except for the relation corresponding to the wavefunction which is a linear combination of the corresponding states. The histories $H_k$ will correspond to parallel universes, and in each of them an observable $\Omega$ will have a value $\omega_k$.

The de Broglie-Bohm interpretation, see \cite{dbb}, appears to be a single universe interpretation, since each particle follows a single trajectory, and all particles are guided by the wavefunction. However, the initial condition on the wavefunction requires an ensemble of the universes with different initial conditions. The question then arises about the status of other histories, i.e. are they also realized in time or are they timeless? The first option would lead to a Many Worlds interpretation without branching, while the second option would mean that a single temporal history is chosen randomly, or picked by some other criterion, like the maximum of the initial probability distribution\footnote{This criterion is also used in the case of the landscape of vacua in String Theory, see \cite{landscp}.}.

\section{Conclusions}

We have presented a Platonic metaphysics where time plays the essential role: it serves to distinguish between real and abstract universes. This role of time together with our proposed mind-brain connection resolves the epistemological problem in platonism \cite{pl}. Namely, if the abstract ideas are outside of spacetime, then how can we do mathematics? According to our approach, the answer is that our mind, which is a temporal sequence of ideas contained in our brain, will contain the copies of these abstract ideas.

An important consequence of the assumption that passage of time is a non-mathematical idea, is that mind is not a mathematical structure. This is because we can imagine processes in time, so that the idea of passage of time is contained in our mind. Hence our mind is not a mathematical structure. This then implies that a mind can not be simulated on a computer.

Note that our approach represents a generalization of von Neumann's idea to formulate QM as a theory of evolving objective universe interacting with human consciousness, see \cite{stap}. In theories with global time, the conflict with relativity can be avoided if one postulates that the label $t$ is not the same as a clock reading and that a clock reading depends on the clock trajectory. On the other hand, it is clear that $t$ is related to the age of the universe. Since $t$ always increases, then the time travel will be impossible within our framework. 

From the platonic perspective, existence of multiple universes is natural, since all possible universes exist as objects in the World of Ideas. Furthermore, by assuming that the passage of time is a fundamental and a non-emergent concept, one can divide all possible universes into the temporal ones and the timeless ones. The temporal universes can be then considered as real, i.e. like the universe we live in, while the timeless universes can be considered as abstract.


\begin{thebibliography}{99}
\footnotesize\bibitem{pl} M. Balaguer, Platonism in metaphysics, Stanford Encyclopedia of Philosophy, (http://plato.stanford.edu/entries/platonism)
\bibitem{teg} M. Tegmark, The mathematical universe, Found. Phys. 38 (2007) 101
\bibitem{penr} R. Penrose, The emperor's new mind: Concerning computers, minds and the laws of physics (Oxford University Press, 1989)
\bibitem{tvar} C.J. Isham, Canonical quantum gravity and the problem of time, in ``Integrable systems, quantum groups, and quantum field theories: proceedings'', ed. L.A. Ibort and M.A. Rodriguez, Nato Advanced Study Institute, Series C: Mathematical and Physical Sciences, vol. 409 (Kluwer, 1993)
\bibitem{ajl} J. Ambjorn, A. Goerlich, J. Jurkiewicz, R. Loll, The nonperturbative quantum de Sitter universe, Phys. Rev. D78(2008) 063544 
\bibitem{mwi} J. Barrett, Everett's relative-state formulation of Quantum Mechanics, Stanford Encyclopedia of Philosophy (http://plato.stanford.edu/entries/qm-everett)
\bibitem{dbb} P.R. Holland,  The quantum theory of motion: An account of the de Broglie-Bohm causal interpretation of Quantum Mechanics (Cambridge University Press, 1993)
\bibitem{landscp} A.N. Schellekens, The emperor's last clothes? (Overlooking the String Theory landscape), arXiv:0807.3249
\bibitem{stap} H.P. Stapp, Quantum theory and the role of mind in nature, Found. Phys. 31 (2001) 1465

\end{thebibliography}
\end{document}